\documentclass[12pt]{article}

\usepackage[a4paper, top=2.5cm, bottom=2.5cm, left=2.5cm, right=2.5cm, centering]{geometry}

\usepackage[style=apa, citestyle=numeric, backend=biber]{biblatex}

\renewbibmacro*{begentry}{\printfield{labelnumber}\labelnamepunct}

\usepackage{csquotes}

\usepackage{graphicx}

\usepackage{hyperref}

\usepackage{authblk}

\usepackage{fancyhdr}

\usepackage{float}

\usepackage[toc]{appendix}

\usepackage[skip=5pt]{caption}
\fancypagestyle{plain}{
  \fancyhf{} 
  \fancyfoot[R]{\thepage}

}

\hypersetup{
    colorlinks,
    linkcolor=blue,
    urlcolor=blue,
    citecolor=blue
}

\begin{document}

\title{\textbf{A network psychometric analysis of maths anxiety factors in Italian psychology students}}

\author[1]{Emma Franchino}
\author[1]{Luciana Ciringione}
\author[2]{Luisa Canal}
\author[2]{Ottavia Marina Epifania}
\author[2]{Luigi Lombardi}
\author[3]{Gianluca Lattanzi}
\author[1]{Massimo Stella}

\affil[1]{CogNosco Lab, Department of Psychology and Cognitive Sciences, University of Trento, Rovereto, Italy.}
\affil[2]{Department of Psychology and Cognitive Sciences, University of Trento, Rovereto, Italy.}
\affil[3]{Department of Physics, University of Trento, Povo, Italy.}
\date{}

\maketitle

\vspace{2em}
\begin{flushleft}
ORCID: \\
Emma Franchino: \href{https://orcid.org/0009-0004-2101-5700}{0009-0004-2101-5700}\\
Luciana Ciringione: \href{https://orcid.org/0000-0002-7430-9022}{0000-0002-7430-9022}\\
Luisa Canal: \href{https://orcid.org/0000-0002-1493-108X} {0000-0002-1493-108X} \\
Ottavia Marina Epifania: \href{https://orcid.org/0000-0001-8552-568X}{0000-0001-8552-568X} \\
Luigi Lombardi: - \\
Gianluca Lattanzi: \href{https://orcid.org/0000-0002-0808-6457}{0000-0002-0808-6457} \\
Massimo Stella: \href{https://orcid.org/0000-0003-1810-9699}{0000-0003-1810-9699} \\
\end{flushleft}
Correspondence concerning this article should be addressed to: \href{mailto:massimo.stella-1@unitn.it}{massimo.stella-1@unitn.it}

\newpage 
\begin{abstract}
  {Dealing with mathematics can induce significant anxiety, strongly affecting psychology students’ academic performance and career prospects. This phenomenon is known as maths anxiety and several scales can measure it. Most scales were created in English and abbreviated versions were translated and validated among Italian populations (e.g. Abbreviated Maths Anxiety Scale). This study translated the 3-factor MAS-UK scale in Italian to produce a new tool, MAS-IT, validated specifically in a sample of Italian undergraduates enrolled in psychology or related BSc programmes. A sample of 324 Italian undergraduates completed the MAS-IT. The data were analysed using Confirmatory Factor Analysis (CFA), testing the original MAS-UK 3-factor model. CFA results revealed that the original MAS-UK 3-factor model did not fit the Italian data. A subsequent Exploratory Graph Analysis (EGA) identified 4 distinct components/factors of maths anxiety detected by MAS-IT. The items relative to “Passive Observation maths anxiety” factor remained stable across the analyses, whereas “Evaluation maths anxiety” and “Everyday/Social maths anxiety” items showed a reduced or poor item stability. Quantitative findings indicated potential cultural or contextual differences in the expression of maths anxiety in today's psychology undergraduates, underlining the need for more appropriate tools to be used among psychology students.}
\end{abstract}

\newpage

\section{Introduction}
Anxiety is an evolutionary feedback mechanism whose goal is to defend oneself from potential dangers and threats \cite{Lovibond1995}. For instance, the thought of meeting a predator might cause physical and emotional discomfort to a tourist visiting a luscious forest. Such discomfort would drive the tourist away without ever physically meeting the predator. In more general terms, anxiety is useful for individuals to avoid threats without ever experiencing dangers. Unfortunately, anxiety can be triggered also by experiences that are not intrinsically dangerous \cite{Hunt2011}. For instance, an equation on a blackboard might trigger feelings of anxiety as strong as the ones inspired by the thought of a voracious predator. Are these thoughts equally threatening? One may think that wild predators are way more dangerous than equations and blackboards. However, for a lot of people also elements related to mathematics can trigger deep sensations of anxiety, and this phenomenon is known as maths anxiety \cite{Ashcraft2007,Stella2022}.

Maths anxiety is included under the generalized anxiety disorders in the Diagnostic and Statistical Manual of Mental Disorders (DSM-V) \cite{Association2013}, however not everyone suffering from maths anxiety would meet the criteria for general anxiety disorder. Furthermore, maths anxiety differs from other anxiety disorders due to its context specificity \cite{Luttenberger2018}. Maths anxiety can be defined as a context-dependent feeling of tension caused by experiences thematically related with mathematics. Individuals might experience maths anxiety when dealing with numbers, when feeling observed by others in maths classes, or when performing tasks mentioning mathematical concepts \cite{Stella2022}. 

The physiological and emotional effects of maths anxiety are quite comparable to the ones of general anxiety (e.g. increased heart rate, stomach ache, feelings of tension, apprehension, nervousness) \cite{Paechter2017, Papousek2012}, whilst some more specific consequences were discovered on the cognitive level. In particular, a study by Ashcraft and Krause (2007) has demonstrated how working memory, maths anxiety and maths performance are strictly related with each other. Working memory is a subcomponent of cognitive memory and it loads/manipulates symbols, rules and skills relative to language/mathematical knowledge \cite{Ashcraft2007}. Among other tasks, working memory intervenes during the solving of mathematical problems by loading information and solving procedural tasks. Working memory has a finite maximum capacity, i.e. it can load a limited maximum amount of information \cite{Beilock2005}. Past findings identified a positive correlation between the complexity of maths problems and the load on working memory \cite{LeFevre2005}. Ashcraft and Krause additionally showed that maths anxiety can drain the resources of working memory, compromising its efficiency and lowering down its maximum capacity \cite{Ashcraft2007}. With fewer cognitive resources available in working memory, individuals affected by maths anxiety can end up getting poorer performances in their maths tasks. This relationship between working memory and maths anxiety can contribute to explaining a negative correlation between maths anxiety and maths achievements \cite{Ashcraft2007}. Moreover, lower performance in solving maths problems can further reinforce feelings of inadequacy or biases \cite{Stella2022}. Interestingly, maths anxiety might co-exist with positive or neutral perceptions of science \cite{stella2020forma,stella2020italian}, making its detection a rather complex and concerning challenge. The most concerning aspect of maths anxiety is that it can cause individuals to avoid mathematics, foster distorted perceptions of the subject \cite{stella2020forma} and perpetuate higher levels of maths anxiety through a vicious cycle \cite{Wilson1999}.

Worldwide statistics have shown that maths anxiety can be found globally in different populations, even in countries known for advanced STEM curricula such as the United States, where 93\% of adults reported experiencing maths anxiety during their education \cite{Blazer2011}. Research has found that students in the humanities and psychology constitute a population deeply affected by maths anxiety \cite{Onwuegbuzie2003}. Psychology students are often required to take several courses in mathematics and statistics throughout their academic studies \cite{Siew2019}. According to Messer and colleagues (1999), nearly 77\% of psychology curricula include at least one statistics-related course, including mathematical content \cite{Messer1999}. For these psychology students, maths-related courses can be particularly challenging and affect academic careers, e.g. delayed graduation, premature career changes, or even failure to complete their degree \cite{Siew2019}. 

Throughout the years, different statistical methods have been created to measure maths anxiety, in order to understand its characterising factors and potential ways to overcome this feelings. Richardson and Suinn (1972) were the first to operationalise the maths anxiety construct by developing the Mathematics Anxiety Rating Scale (MARS, \cite{Richardson1972}). The MARS, designed to measure maths anxiety in various contexts, presented a test-retest reliability of .85 and a Cronbach’s alpha of .97 \cite{Capraro2001}. Analysing the MARS, Rounds and Hendel (1980) identified two main factors that were being assessed: anxiety related to the evaluation and learning of mathematical skills, and anxiety associated with everyday situations employing maths abilities \cite{Rounds1980}. Later, Resnick and colleagues (1982) detected a third factor, focusing on mathematical skills required in social contexts, e.g. interactions with peers and teachers \cite{Resnick1982}. 

Although MARS has been one of the most cited scales for measuring maths anxiety, its length, consisting of 98 items, prompted researchers to create more user-friendly versions. A revised and shorter version of MARS, known as MARS-R, was created by Hopko (2003), in which 12 items were removed yielding a two-factor model \cite{Hopko2003confirmatory}. 

One of the latest developed maths anxiety scales is the Mathematics Anxiety Scale-UK (MAS; \cite{Hunt2011}), which was originally validated in a British undergraduate population. MAS-UK includes 23 items related to statements about situations involving mathematics and maths anxiety (MA). The items delineate 3 main factors contributing to maths anxiety:

\begin{enumerate}
    \item Evaluation MA, which refers to anxiety related to assessing one's mathematical abilities, often in formal academic settings (e.g. taking a maths exam or answering questions in class).
    \item Everyday/Social MA, elicited in daily situations where maths is required, often with social implications (e.g. calculating change, splitting bills, or remembering phone numbers).
    \item Passive Observation MA, experienced when passively observing maths-related activities without direct involvement (e.g. watching someone solve a problem or listening to a maths lecture).
\end{enumerate}

To the best of our knowledge, within the Italian landscape, the main MA psychometric scale validated among Italian participants is the Abbreviated Maths Anxiety Scale (AMAS, \cite{Hopko2003}). This scale consists of 9 items, measuring Learning Maths Anxiety (LMA) and Maths Evaluation Anxiety (MEA), and was found to be more parsimonious (i.e. briefer and externally valid) than the MARS-R scale \cite{Hopko2003}. Importantly, despite featuring “abbreviated” in its name, AMAS is not a subsample of items from MAS-UK or other scales but rather a substantial rewriting of maths anxiety items all relative to learning and testing in mathematical contexts. These elements provide a substantial difference between AMAS and MAS-UK and other psychometric tools.

Relative to Italian samples, AMAS was firstly translated into Italian and tested for its validity and reliability among high school and college students \cite{Primi2014}. Afterwards, AMAS was also validated among Italian primary school children by Caviola and colleagues in 2017 \cite{Caviola2017}. The Italian translation of AMAS was then used by different studies in Italian population samples: Piccirilli and colleagues (2023) sampled 73 high school students \cite{Piccirilli2023}; Cuder (2024) investigated 109 middle school students \cite{Cuder2024}. Additionally, a new adapted version of AMAS was developed by Prima and colleagues in 2020 in order to measure maths anxiety in children: the Early Elementary School Abbreviated Maths Anxiety Scale (EES-AMAS, with 9 items; \cite{Primi2020}).

Although AMAS is the primary scale used in Italy to measure maths anxiety, the present study focuses on the MAS-UK scale because it adopts a 3-factor model that aligns with other psychological insights from earlier scales like MARS-R and MARS. Specifically, MAS-UK is more advantageous than AMAS because it not only considers the Evaluation and Passive elements of MA, which can be captured also by the 2-factor model of AMAS, but also includes a range of everyday/social situations in which maths anxiety might be relevant. In contrast, AMAS does not have a specific factor for social contexts; its structure primarily revolves around evaluating mathematical tasks and learning maths content. In agreement with other studies \cite{Stella2022,Devine2018,Luttenberger2018} emphasizing the relevance of social relationships in the diffusion of maths anxiety (e.g. parents transmitting maths anxiety to their children, peer evaluation exacerbating maths anxiety in teenagers), we have focused on a psychometric tool that adequately addresses the social factor. In light of these considerations, the aim of the present study is to validate an Italian-translated MAS-UK scale - from now on referred to as MAS-IT - among a sample of Italian undergraduate students in psychology or related fields. 

Additionally, considering that MAS-UK was created in 2011, some of its items might no longer be optimal for measuring maths anxiety among today's students. Indeed, many cultural changes could have influenced academic contexts, such as technological innovations like social media, the COVID-19 pandemic and other societal changes. Hence, a secondary goal of the current study is to evaluate how the MAS-IT behaves in a sample of Generation-Z individuals. To this aim, network psychometrics \cite{Golino2017} was used to assess the structure of correlations among maths anxiety items in the MAS-IT responses coming from Gen-Z psychology students in Italy.

\section{Methods}
\subsection{Participants} 
A sample of 324 students from the Department of Psychology in an Italian University were recruited through convenience sampling via the Department’s social media platforms and word of mouth. Students were enrolled in different Bachelor’s programs related to psychology: Cognitive Psychology Sciences and Techniques and Communication Interfaces and Technologies. The participants were all adults (above 18 years old) and native Italian speakers. The study protocol was approved by the Human Research Ethics Committee of the Local Ethical Committee (ID: 2024-039) and adhered to the principles of the Declaration of Helsinki. Participants provided informed consent and received no compensation for their involvement.

The sample of participants consisted of different groups, based on the year they were recruited, and the academic year they were enrolled in at that time. The present study was conducted both in 2023 and 2024. In 2023, students were not grouped by academic enrolment year, because of numerosity issues, resulting in a heterogeneous larger group. In 2024, the sample was divided into first, second and third academic years (which correspond to the subdivision of the Italian Bachelor’s degree system).

\subsection{Materials}
\subsubsection{MAS-IT scale}
For this study, an Italian translated version of the MAS-UK was used, which will be referred to as MAS-IT, in order to be consistent with the original name. The MAS-IT consisted of 23 different items, corresponding to the retained items in the final version of the MAS-UK, which were carefully translated into Italian by two experts, cf. Table \ref{tab:table_items}. The translated items corresponded to statements on situations related to mathematics; such statements could be grouped into 3 factors, depending on which aspects of mathematics - and MA - they measured (i.e. the same factors identified by MAS-UK; \cite{Hunt2011}). These factors, which have been previously mentioned in the Introduction, correspond to: Evaluation MA, Everyday/Social MA and Passive Observation MA. Both the translation of the items from the MAS-UK and their grouping in different factors can be found in Table \ref{tab:table_items}.

To reduce semantic wording effects and given the absence of requirements for semantic coherence in the original MAS-UK scale (cf. \cite{Hunt2011}), MAS-IT items were randomised in their order of presentation. Participants were asked to rate their anxiety, relative to each item, on a 5-point Likert scale, ranging from 1 (not anxious at all) to 5 (very anxious), providing a broad but approximate measure of their maths anxiety levels. The answers were collected in an Excel file. 

\subsection{Data Analysis}
Descriptive analyses (Table \ref{tab:stats_labels}) were conducted using Python. The same programming language was employed for comparative analyses based on factors and students’ year of enrolment, as well as for generating representative box plots. In instances where the sample included fewer than 50 students per enrolment year, both non-parametric and parametric analyses were utilised, with their convergent results being assessed despite sample size limitations. For non-parametric analyses, Kruskal-Wallis (\textit{KW}) and Dunn-Bonferroni tests were performed, while ANOVA was used for parametric analyses. These methods were adopted to explore differences between enrolment years within the sample. To facilitate comparison, the mean Likert score of students in each population group, as well as the mean score for all items within each MA factor, was calculated, providing a clearer representation of the comparisons of interest.

Moreover, a correlation matrix (using Kendall Tau rank correlations) for each item of MAS-IT was calculated and the respective correlogram was generated. The correlogram illustrates the strength and direction of the correlations, with items visually grouped according to their respective factors. 

\subsubsection{Confirmatory Factor Analysis}
Further analyses, including Confirmatory Factor Analysis (CFA) and Exploratory Graph Analysis (EGA), were performed using the \texttt{R}-packages \texttt{lavaan} \cite{Rosseel2012} and \texttt{EGAnet} \cite{Golino2017}, respectively. The goal was to assess the reliability of MAS-IT within this specific population sample. Precisely, the Confirmatory Factor Analysis was conducted to evaluate whether the hypothesized 3-factor structure of the MAS-UK would fit the observed data in the sample of the Italian psychology students~\cite{Brown2015}. 

\subsubsection{Exploratory Graph Analysis}
Exploratory Graph Analysis \cite{Golino2017} was used to uncover the latent factor structure of the MAS-UK items based on the collected MAS-IT data. Unlike CFA, which tests a predefined model structure, EGA identifies factor structures directly from the data using network psychometrics \cite{Christensen2021}. EGA involves three steps: 

\begin{itemize}
    \item {\bf Correlation Matrix Estimation}. This step involves using the \texttt{cor\_auto} method to estimate correlations based on data type, allowing the evaluation of correlations between all pairs of variables. 
    \item {\bf Graphical Least Absolute Shrinkage and Selection Operator (GLASSO)}. The GLASSO algorithm is applied to the correlation matrix to estimate a sparse partial correlation network, preserving stronger connections over weaker ones. By penalising weaker correlations, GLASSO effectively identifies the strongest connections between items, resulting in a clear and interpretable network structure.
    \item {\bf Community Detection and Factor Estimation}. This step groups items into clusters (i.e., factors) based on their network connections. These clusters represent latent constructs, where items within the same cluster share strong associations \cite{Golino2017}. 
\end{itemize}

Item stability analysis was performed using the bootstrap EGA (with the \texttt{bootEGA} function, provided by \texttt{EGAnet}), to estimate and evaluate the stability of dimensions identified by exploratory graph analysis, as well as to assess the robustness of each item’s placement within those dimensions \cite{Christensen2021}.

\subsubsection{Correlation between EGA and MAS-UK factors}
Once the MAS-IT items were clustered into different communities/factors by EGA, a comparison was made with the factor structure delineated by MAS-UK. The Jaccard similarity Index was calculated based on the intersection and union of items within each factor, providing a measure of how similar the EGA and MAS-UK factors are (cf. \cite{Stanghellini2024}). A Jaccard index of 1 indicates perfect similarity between two factors, while an index of 0 indicates complete dissimilarity, meaning that the factors do not share any items.

Next, all item scores within each factor were summed, considering both the CFA factors (equivalent to the MAS-UK factor structure) and the \texttt{EGAnet} factors. Kendall Tau correlations were then computed between pairs of factor scores. These correlations are presented as a correlogram, highlighting the strength and direction of the correlations between the EGA communities and the MAS-UK factors.

\section{Results}
Table \ref{tab:stats_labels} reports a summary of the main descriptive statistics of the collected data. These values suggest a potential lack of normality in the distributions of scores for individual items. Combined with the sample size limitations (see Methods), this highlights the need to compare parametric and non-parametric approaches when detecting differences across subsamples.

Further comparative analysis were conducted, focusing primarily on differences between population groups. Specifically, the analysis considered  only the division of the sample based on academic enrolment year (First year: ${N_{1}}$ = 48; Second year: ${N_{2}}$ = 68; Third year: ${N_{3}}$ = 51), i.e. the data from students who participated in the study in 2023 were excluded, as a clear distinction between academic years was required. 

\begin{figure}[H]
    \includegraphics[width=0.95\linewidth]{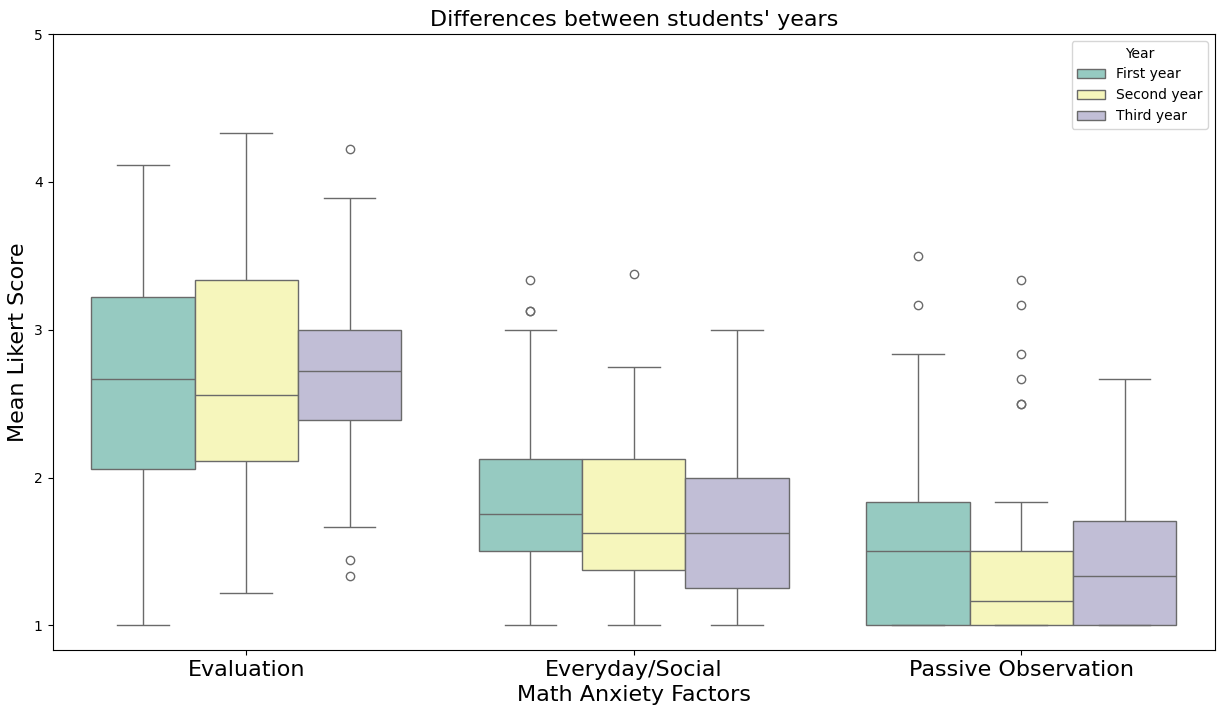}
    \caption{Box plot comparing mean Likert scores of MAS-IT questions, grouped by MAS-UK factors, over students’ academic years. Empty dots indicate outliers; bars indicate data distribution.}
    \label{fig:comparison_population}
\end{figure}

The data were also grouped by Maths Anxiety factor, providing a clear contrast between sample groups. The comparative box plot (Figure \ref{fig:comparison_population}), where each box represents the mean Likert score of all students in each sample group, indicates that no significant differences were found between student groups for any of the MA factors (Evaluation MA: \textit{KW statistics} = 0.08, \textit{p} = .959; Everyday/Social MA: \textit{KW} = 2.14, \textit{p} = .343, Passive Observation MA: \textit{KW} = 4.3, \textit{p} = .117). While the boxes show varying mean Likert scores within each MA factor, the scores remain consistent across the sample groups.

\begin{figure}[H]
    \centering
    \includegraphics[width=1\linewidth]{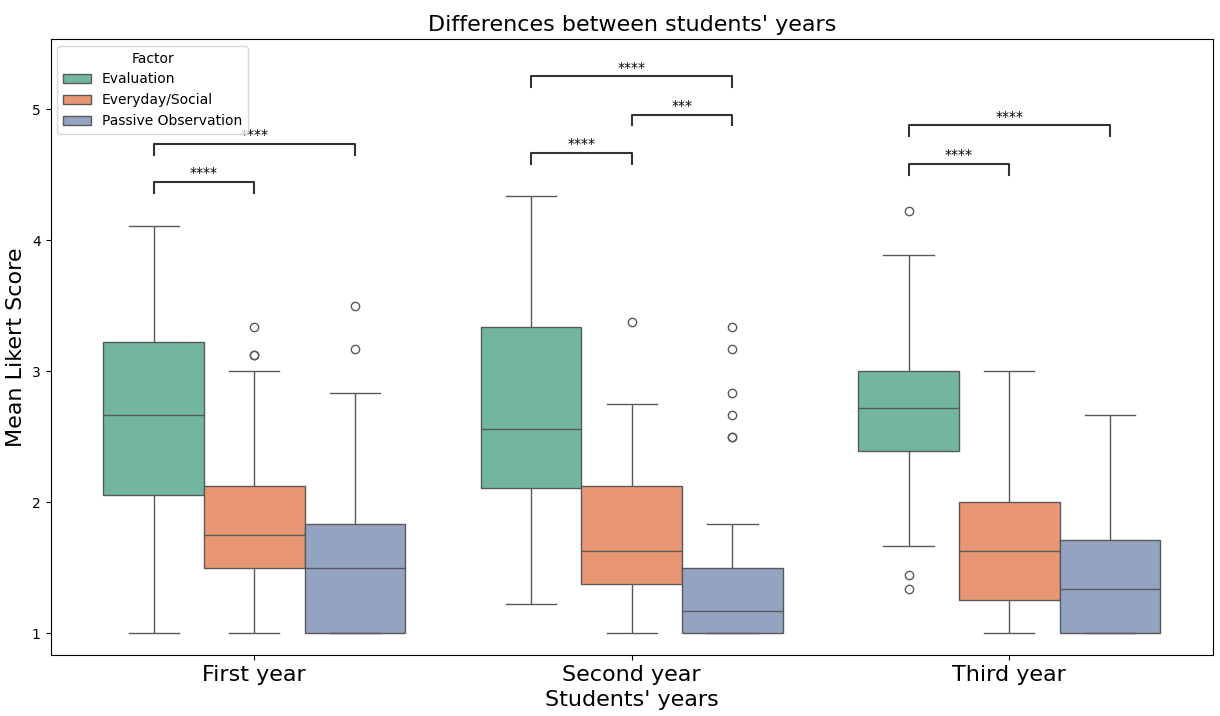}
    \caption{Box plot comparing mean Likert scores of MAS-IT questions over MAS-UK factors, grouped by students’ academic years. Empty dots indicate outliers; bars indicate data distribution.}
    \label{fig:factors_comparison}
\end{figure}

On the other hand, the comparison between MA factors, grouped by student cohorts, reveals significant differences (\textit{p} $<$ .001), as shown in Figure \ref{fig:factors_comparison}. Specifically, across all academic enrolment years, significant differences are evident between Evaluation MA and Passive Observation MA, as well as between Evaluation MA and Social MA. Additionally, among students enrolled in the second year of their BSc program, a significant difference was found between Social MA and Passive Observation MA. These findings indicate that the interviewed students exhibited varying levels of maths anxiety across different factors. Figure \ref{fig:factors_comparison} illustrates that levels of Evaluation MA are higher than those of Social MA or Passive/Observation MA. This suggests that psychology undergraduate students may experience higher levels of maths anxiety when performing specific maths evaluation tasks, i.e. solving an equation.

\subsection{Correlational Analysis}
\begin{figure}[H]
    \centering
    \includegraphics[width=0.95\linewidth]{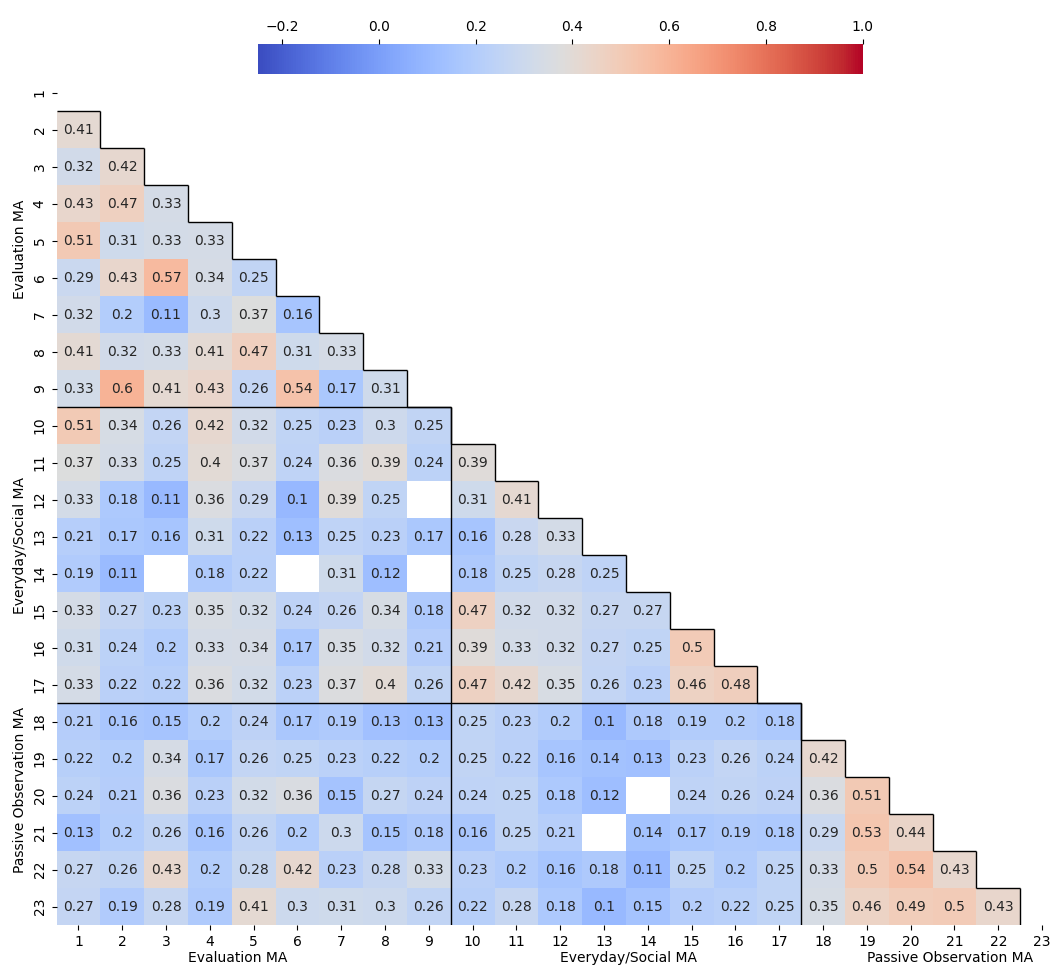}
    \caption{Correlogram of MAS-IT items grouped by MAS-UK factors. Blank spaces represent non significant correlations (\textit{p} $\geq$ .05) between items.}
    \label{fig:correlogram}
\end{figure}

Figure \ref{fig:correlogram} presents the correlations between each MAS-IT item. The strongest positive correlations (i.e. $\tau$ coefficient close to 1, represented by red squares) were observed between items within the same MA factor. Notably, no negative correlations were identified between MAS-IT items, and six non-significant correlations (\textit{p} $\geq$ .05) emerged.Of these, four occurred between items from the Everyday/Social MA factor and the Evaluation MA factor, while the remaining two involved correlations between the Passive Observation MA and the Everyday/Social MA factors.

Higher correlations between items indicate that total scores are more consistent across the two allocations of items within factors/dimensions. Since the current results show stronger correlations between items within the same factors, the 3-factor model appears to be a coherent approach for the division of items. 

\subsection{Confirmatory Factor Analysis with the UK 3-factor model}
The Confirmatory Factor Analysis (CFA) tested the hypothesized original factor structure of the MAS-UK against the collected data. The model's goodness of fit was evaluated using standard indices, i.e. the Comparative Fit Index (CFI) $>$ 0.90, the Root Mean Square Error of Approximation (RMSEA) $<$ 0.06, and the Standardized Root Mean Square Residual (SRMR) $<$ 0.08 \cite{Hu1999}.

The CFA performed on the MAS-IT data revealed a significant discrepancy between the original model and the MAS-IT data, leading to the acceptance of the alternative hypothesis, which posits that the original 3-factor MAS-UK model does not fit the data ($\chi^2$ = 898.803, \textit{df} = 227, \textit{p} $<$ .001). The fit indices further supported this finding: the RMSEA value was 0.096, indicating an inadequate fit, and the CFI, which compares the hypothesized model to a null model, was 0.811, below the acceptable threshold of 0.90. Conversely, the SRMS index suggests a satisfactory fit (0.079 $<$ 0.08). Overall, these results indicate that the UK 3-factor model does not fully align with the Italian data, highlighting the need to refine the model or make theoretical adjustments to better capture the underlying structure of the data. To address this, the analysis was extended through exploratory approaches. 

\subsection{Exploratory Graph Analysis}
Given the inadequacy of the UK model for the Italian data, exploratory analyses were required. To address this, an Exploratory Graph Analysis (EGA) was performed to cluster items into new factors (see Methods). Subsequently, a Unique Variable Analysis (UVA) \cite{Christensen2023} was conducted to identify potentially redundant items by calculating the weighted redundancy index (wTO). This approach allowed us to determine which item pairs could be considered locally dependent, or redundant. Finally, an item stability analysis was carried out \cite{Christensen2021}: the satbility of the item communities, previously identified in the median graph, was assessed through bootstrap analysis.

\subsubsection{Network Estimation and Local Network Properties}
\begin{figure}[H]
    \centering
\includegraphics[width=0.6\linewidth]{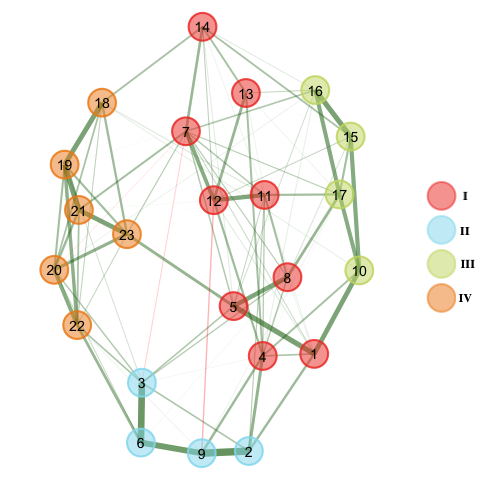}
    \caption{MAS network representing the clustering of the MAS-IT items and the strength of correlations between them.}
    \label{fig:mas_network}
\end{figure}

EGA identified the different components or factors of maths anxiety using the MAS-IT collected data, comparing them with the original MAS-UK model. The network analysis estimated four clusters (also referred to as factors or communities), as shown in Figure \ref{fig:mas_network}, capturing various dimensions of maths anxiety.

Factor I includes the largest number of items, all related to situations where individuals need to use their maths abilities to solve a maths problem. These situations may involve a social context (e.g. item 4: “Being asked to calculate € 9.36 divided by four in front of several people”), or can be related to the evaluation of one’s maths skills (e.g. item 8: “Being asked to calculate the three fifths of a percentage”). Factor II comprises only four items, all specifically related to maths evaluation. These scenarios may occur in a classroom setting (e.g. item 2 “Being asked to answer a question on the board in front of the class”), or involve formal assessments (e.g. item 3 “Taking a maths exam”). Factor III also includes four items, focusing on social or everyday maths anxiety (e.g. item 17 “Working out how much your shopping bill comes to”). Finally, Factor IV is the only one that perfectly aligns with one of the original MAS-UK's factors, including all the items within the Passive Observation MA factor.

To further analyse the similarity between the clustering of items into separate communities by the EGA and the factors delineated by the MAS-UK, the Jaccard similarity index was calculated. The results indicated that the EGA Factor I, as well as Factor II, are most similar to the Evaluation MA factor (delineated by the MAS-UK),with Jaccard similarity indices of 0.38 and 0.44, respectively. On the other hand, Factor III aligns most closely with the Everyday/Social MA factor, showing a Jaccard similarity index of 0.5. Finally, as previously mentioned,Factor IV perfectly matches the Passive Observation MA factor, resulting in a Jaccard index of 1 between the two factors.

\begin{figure}[H]
    \centering
    \includegraphics[width=0.6\linewidth]{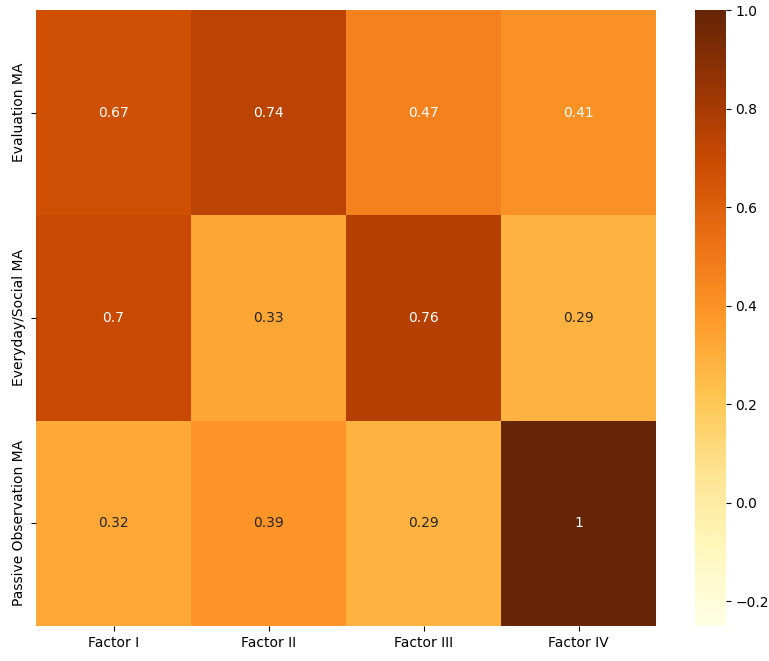}
    \caption{Correlogram between EGA factors and MAS-UK factors on MAS-IT data. All correlation coefficients (Kendall-Tau) resulted significant (\textit{p} $<$ .05).}
    \label{fig:correlogram_ega_mas}
\end{figure}

The statistical correlation between the sum of the scores for all items within each EGA factor and each MAS-UK factor was calculated using the Kendall Tau coefficient and visualized in a correlogram (Figure \ref{fig:correlogram_ega_mas}). The correlogram reveals that no negative correlations were present and that the strongest correlations ($\tau$ $>$ 0.6) occurred between the factors identified as the most similar by the Jaccard index. The only exception is EGA Factor I, which shows a high correlation not only with the Evaluation MA but also with Everyday/Social MA. This finding further suggests that the EGA Factor I includes items combining both Evaluation MA and Everyday/Social MA.

\subsubsection{Redundancy between items}
The Unique Variable Analysis (UVA) on the MAS-IT, based on the weighted redundancy index (wTO), was conducted with \texttt{EGAnet}. The analysis identified  several pairs of items with significant redundancy (wTO $>$ 0.20). The highest redundancy (wTO = 0.308) was observed between items 2 and 9 (cf. Table \ref{tab:table_items} in the Appendix), both of which pertain to situations where participants are asked to solve maths problems in front of a class. A moderate level of redundancy was also found between items 3 and 6, which are related to maths exams. Additionally, small-to-moderate redundancy (wTO between 0.20 and 0.25) was detected across five different item pairs. The highest redundancy within this range appeared between items 19 and 21, both describing passive circumstances, and between items 6 and 9, which involve maths examinations in a classroom setting. In contrast, lower redundancy was noted between items 15 and 16 (wTO = 0.214), which describe everyday situations involving monetary calculations, and between items 18 and 19 (wTO = 0.212), both referring to listening to maths-related talks. Finally, the lowest redundancy (wTO = 0.205) was identified between items 21 and 23, both representing passive observation of someone solving a maths problem. 

The items with the highest redundancy are those related to the most similar circumstances, indicating that they measure the same aspect of MA. Therefore, in future studies aimed at developing a new Maths Anxiety scale, retaining only one item from each highly redundant pair could be a viable strategy. As recommended by Christensen and colleagues (2023), a conservative approach to merging redundant items involves considering only those with a wTO exceeding the 0.3 threshold \cite{Christensen2023}. In this study, this criterion applies solely to items 2 and 9. As noted in the Introduction, a shorter MA scale, the AMAS, has previously been employed with Italian samples and demonstrated validity. However, this scale has not yet been validated specifically within a population of psychology students. Considering these insights, future studies could aim to develop and validate a shorter, more optimal MA scale tailored to this undergraduate population. 

\subsubsection{Item stability analysis}
\begin{figure}[h]
    \centering
    \includegraphics[width=0.7\linewidth]{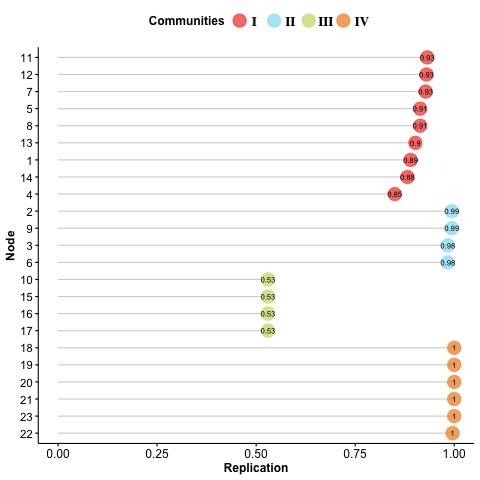}
    \caption{Item stability analysis for the MAS-UK scale in the sample of Italian psychology students.}
    \label{fig:mas_stability}
\end{figure}

Figure \ref{fig:mas_stability} presents the results of the item stability analysis, conducted using the \texttt{bootEGA} function from the \texttt{EGAnet} package. The y-axis lists each item of the MAS-IT questionnaire, organized according to the four distinct communities (or factors) identified by the previous Network Estimation analysis. These factors are visually differentiated by colour, matching those used in the network estimation plot (Figure \ref{fig:mas_network}). The x-axis represents the stability level of each item after 500 bootstrap replications, with values ranging from 0 (lowest stability) to 1 (highest stability). 

The stability score (\textit{S}) is higher than 80\% for most items. At the factor level, items within Factor IV, which correspond to the Passive Observation MA factor in the MAS-UK, demonstrated perfect stability with a score of 100\%. Slightly lower stability scores were observed for Factor II's items (\textit{S} = 0.98\% and \textit{S} = 0.99\%), while Factor I's items showed stability scores ranging from 0.85\% to 0.93\%. Conversely, the lowest stability score (\textit{S} = 0.53\%) was found for all items within Factor III.

These results demonstrate that, through non-parametric bootstrapping, which repeatedly resamples the data with replacement, some items within the EGA factors consistently cluster within the same factor, as indicated by  higher stability scores. Conversely, Factor III exhibited low stability, with its items being inconsistently assigned across bootstrap samples. This instability may suggest ambiguity in the content of these items. As shown in Table \ref{tab:table_items}, these items all relate to the use of maths skills for calculating expenses, presumably involving cash transactions. Given this observation, further investigation into the underlying causes of the low stability is desirable. 

The bootstrap test for the Total Entropy Fit Index (TEFI) was then performed to assess whether the four-community structure identified by EGA provided a better fit than a random 4-dimensional structure. The results showed that the Base TEFI value (M = -18.31, SD = 0.46) was significantly lower than the Comparison TEFI (M = -14.63, SD = 0.26), (\textit{p} = .002, one-tailed). This finding confirms that a random allocation of the MAS-IT items into factors does not adequately explain the 4-factor structure delineated by EGA. Therefore, despite the instability of some items (specifically those in Factor III), the overall 4-factor structure remains a mostly good fit for the MAS-IT data.

\section{Discussion}
Mathematics anxiety is a well-documented psychological phenomenon with significant implications for academic performance \cite{Stella2022}, particularly in disciplines that require quantitative skills \cite{Siew2019,Primi2020}. This study examined mathematics anxiety factor scores across different enrolment years within a Psychology BSc programme. Our results provide insights into the stability of maths anxiety levels over time and highlight the relative prominence of specific anxiety dimensions as defined by the MAS-UK scale \cite{Hunt2011}, which was translated into Italian for this study and referred to as MAS-IT.

It is important to note that the present work differs substantially from the validation of the AMAS in various Italian populations \cite{Primi2014,Caviola2017}, as the AMAS and MAS-UK/MAS-IT contain different items and, therefore, measure different aspects of MA \cite{Hopko2003}. While the AMAS is a shorter scale that assesses the same broad psychological construct - maths anxiety - it remains fundamentally distinct from the MAS-UK and, by extension, the MAS-IT. 

Relevant literature provides substantial evidence supporting the need to study maths anxiety among psychology students. For example, a study by Paechter and colleagues (2017) found that psychology undergraduates often experience significant maths and statistics anxiety, which negatively affects their learning behaviours and exam performance \cite{Paechter2017}. Similarly, research by Wilson (1999) suggests that students with high maths anxiety tend to avoid enrolling in courses with substantial quantitative content, thereby limiting their exposure to and proficiency in essential research skills \cite{Wilson1999}. Given the results of this study and the existing literature, addressing maths anxiety within psychology curricula is crucial to improving students' engagement and competence in quantitative aspects of their education.

A key finding of the present study is that mathematics anxiety levels remained stable across enrolment years. This suggests that, within this sample, students entering the psychology programme exhibit similar levels of maths anxiety, despite receiving more statistics courses and training as psychologists. Given that maths anxiety can stem from both early educational experiences and broader societal influences \cite{Ashcraft2002,Stella2022}, the consistency observed here may reflect two possibilities: (i) a lack of systematic change in mathematics education or (ii) a lack of anxiety-reducing initiatives when students enter higher education. This finding also suggests that, despite potential variations in teaching methodologies or curriculum changes over time, core anxieties related to mathematics persist among psychology students, who often self-select into the discipline \cite{Wilson1999} due to its perceived lower mathematical demands compared to STEM fields. 

In addition to its stability across years, maths anxiety was not experienced uniformly across different factors. Specifically, students exhibited higher scores on the Evaluation Maths Anxiety factor compared to the Everyday/Social Maths Anxiety and Passive Observation Maths Anxiety factors. This finding aligns with both theoretical \cite{Richardson1972,Maloney2012} and empirical literature \cite{Cipora2022} on maths anxiety, which suggests that evaluative contexts - such as examinations, graded assignments, or situations requiring public demonstrations of mathematical ability - provoke the highest levels of anxiety.

Another important finding of the current study is that the 3-factor model proposed by the MAS-UK scale \cite{Hunt2011} did not fit the MAS-IT data. This result led to further exploratory analysis, conducted through network psychometrics \cite{Golino2017,Christensen2021, Stella2022,Christensen2023}. This approach revealed four factors, which fit the data better than random expectation. 

Regarding the stability of MAS-IT items, the lowest stability scores were found for the factor containing items related to everyday situations where individuals are asked to use maths to deal with expenses and money, particularly cash (see items 10, 15, 16, 17 in Table \ref{tab:table_items}). 
Considering that the MAS-IT data was collected in 2023 and 2024 and that the MAS-UK questionnaire was created in 2011, it is plausible that the rise of cashless payments in recent years - especially among students - contributed to the low stability of these items. We suggest that digital payment methods allow users to avoid tasks that typically require maths skills, such as splitting bills or calculating change. As demonstrated in a study by Paundra and colleagues (2023), Millennials and Generation-Z have increasingly shifted to digital paying methods, with up to 69\% of respondents from these generations reporting using digital money more often than cash \cite{Paundra2023}. This shift in payment habits, driven by cultural changes, might explain the lower stability of these items. Importantly, this instability is unlikely to be due to a small sample size, as network psychometrics was able to reproduce the factors of the original MAS-UK scale.

The present study has a few limitations. Firstly, our sample size was not sufficient to perform a CFA of the 4-factor model identified in this study. Through future data collection, we aim to conduct a CFA, excluding unstable items and developing a refined MAS-IT scale. Secondly, our approach did not assess the stability of the MA scale within the same participants as they progressed through their curricula. This limitation could be addressed by collecting larger longitudinal datasets that follow psychology students throughout their academic careers, as in a Lagrangian approach for fluid dynamics. Thirdly, it is important to highlight that maths anxiety is a complex phenomenon \cite{Stella2022}, which may not be fully explained by psychological factors alone. Instead, it could be reduced \cite{stella2020forma} or influenced \cite{Cipora2022} through social interactions. Future studies could explore maths anxiety with richer behavioural data.

Considering our current results, future studies could implement an alternative scale to measure MA in Italian Psychology undergraduates. This new scale could be made shorter than the MAS-UK by removing redundant and unstable items that did not provide significant information. Additionally, new items could be developed to better reflect the cultural context and the specific experiences of the target population. An innovative approach could involve AI models - such as \textit{AI-GENIE} (cf. \cite{Russell2024}) - in order to select the most appropriate items for capturing the distinct constructs of maths anxiety and to support the development of a robust new psychometric tool.

In view of this future development and potential systematic reviews on the topic, we here release the dataset of responses for the MAS-IT scale: \href{https://doi.org/10.17605/OSF.IO/3KYDQ}{MAS-IT dataset}.

\section{Conclusion}

Studying maths anxiety through psychometric methods is crucial, given its widespread prevalence, especially among psychology undergraduates, the focus of this study. Our findings demonstrated the inadequate fit of the 3-factor UK model to the data collected from Italian psychology and reveal the low stability of certain items. 

This study contributes valuable data to the scientific community, supporting the development of robust psychometrics tools tailored to this specific population. Furthermore our results emphasize the critical need to establish effective support strategies to help students manage (if not overcome) maths anxiety. Addressing this issue could significantly enhance their academic experience and improve their overall performance. 

\newpage

\section*{Declarations}
\textbf{Funding} No funds, grants, or other support were received during the preparation of this manuscript.\\
\\
\textbf{Ethics approval} The study protocol was approved by the Human Research Ethics Committee of the Local Ethical Committee (ID: 2024-039) and adhered to the principles of the Declaration of Helsinki.\\
\\
\textbf{Consent} Informed consent was obtained from all individual participants included in the study.\\
\\
\textbf{Conflict of interest} The authors have no relevant financial or non-financial interests.\\

\newpage
\printbibliography

\newpage
\begin{appendices}
\section*{Appendix}
\begin{table}[H]
    \centering
    \includegraphics[width=0.95\linewidth]{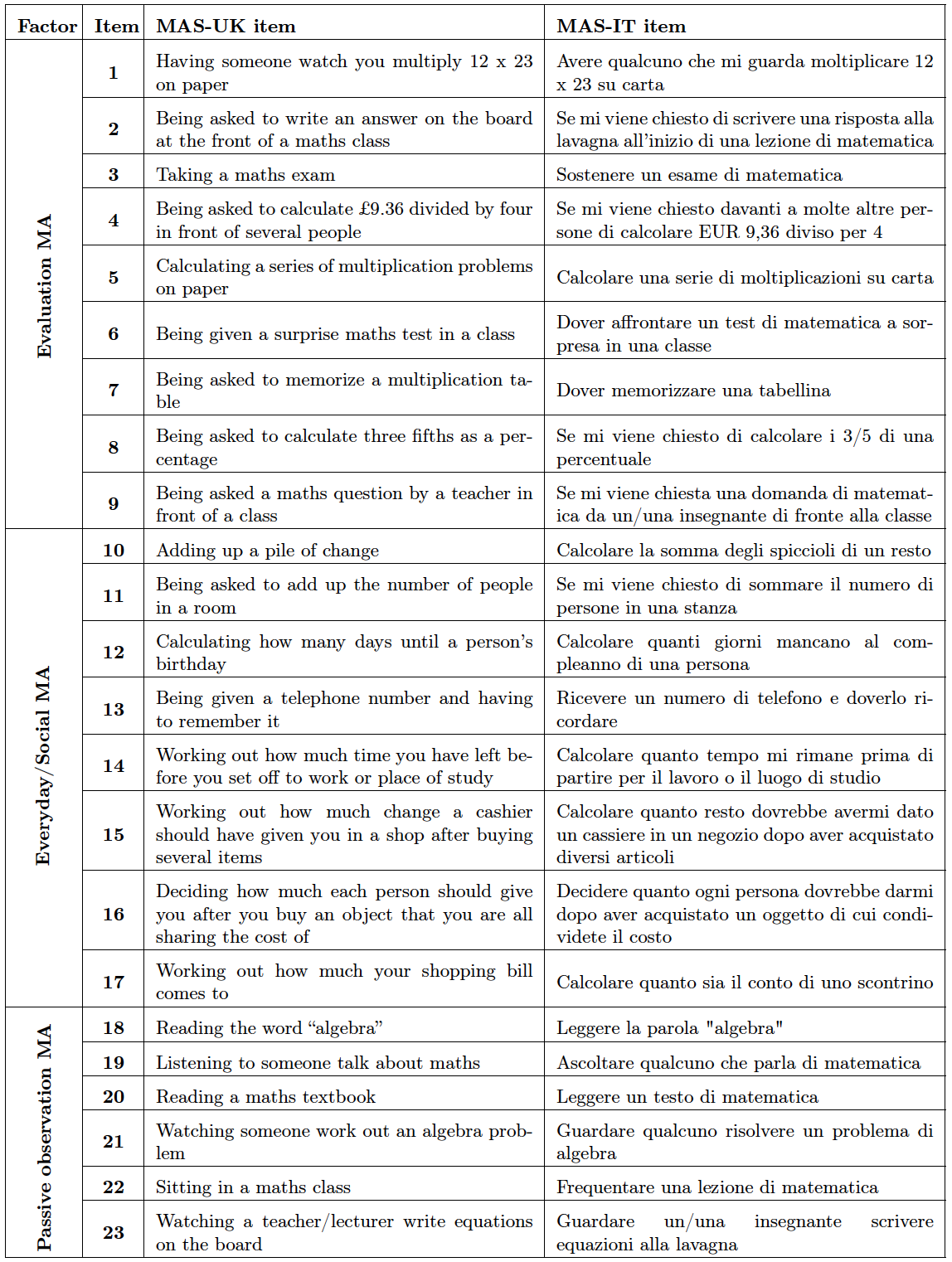}
    \caption{MAS-UK items and correspondent MAS-IT items.}
    \label{tab:table_items}
\end{table}

\begin{table}[ht]
    \centering
    \includegraphics[width=0.85\linewidth]{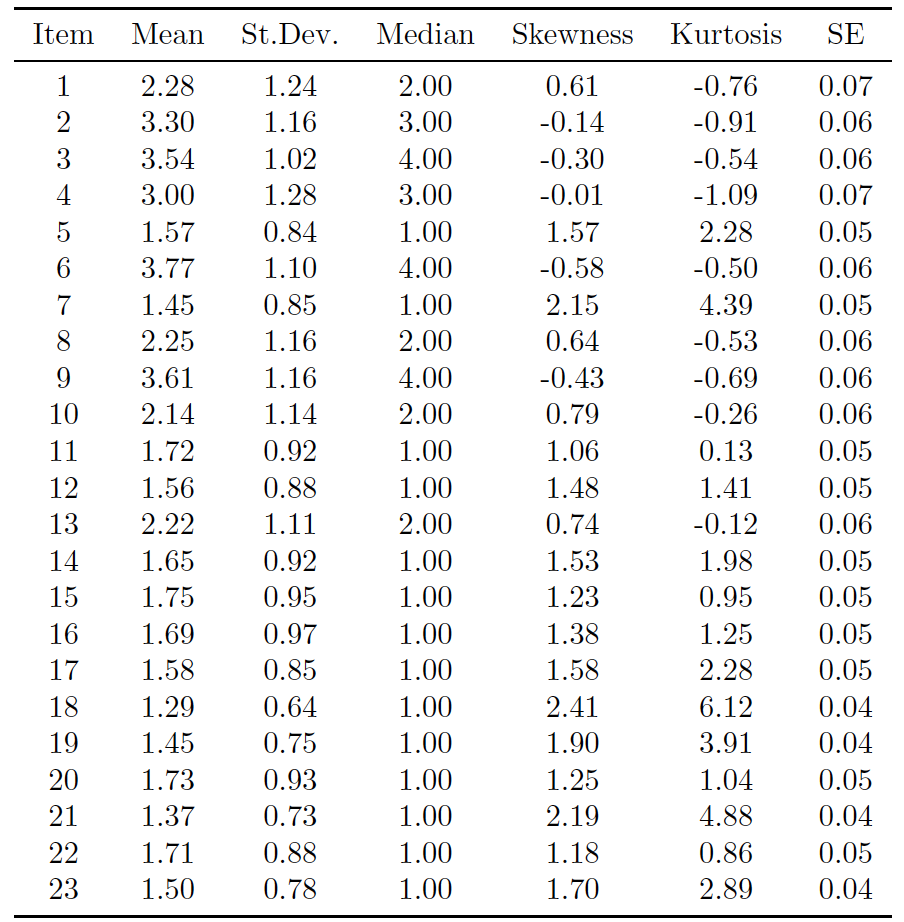}
    \caption{Descriptive statistics of the MAS-IT items. The minimum and maximum values assignable to each item were the lowest and highest point of the Likert scale: respectively 1 and 5.}
    \label{tab:stats_labels}
\end{table}
\end{appendices}

\end{document}